\documentclass
[aps,prd,superscriptaddress,preprint,preprintnumbers,tightenlines,nofootinbib,floatfix]{revtex4-1}
\def \brf{{\cal B}}
\def \bea{\begin{eqnarray}}
\def \beq{\begin{equation}}

\def \eea{\end{eqnarray}}
\def \eeq{\end{equation}}
\def \egh{\rm{E}_\gamma^{\rm high}}
\def \egl{\rm{E}_\gamma^{\rm low}}
\def \bea{\begin{eqnarray}}
\def \beq{\begin{equation}}

\def \eea{\end{eqnarray}}
\def \eeq{\end{equation}}

\newcommand{\electron}{\mbox{$e^{-}$}}

\newcommand{\eminus}{\electron}
\newcommand{\positron}{\mbox{$e^{+}$}}
\newcommand{\eplus}{\positron}

\newcommand{\ee}{\eplus\eminus}                 
\newcommand{\muplus}{\mbox{$\mu^{+}$}}
\newcommand{\muminus}{\mbox{$\mu^{-}$}}
\newcommand{\mumu}{\muplus\muminus}
\newcommand{\goesto}{\mbox{$\rightarrow$}}

\newcommand{\upsi}{\mbox{$\Upsilon$}{\rm (1S)}}
\newcommand{\upsii}{\mbox{$\Upsilon$}{\rm (2S)}}
\newcommand{\upsiii}{\mbox{$\Upsilon$}{\rm (3S)}}

\newcommand{\piz}{\mbox{$\pi$}^{0}}

\newcommand{\pizpiz}{\mbox{$\pi$}^{0}\mbox{$\pi$}^{0}}
\newcommand{\gamgam}{\gamma\gamma}
\newcommand{\chibzero}{\mbox{$\chi_{b0}(1P)$}}

\newcommand{\chibj}{\mbox{$\chi_{bJ}(1P)$}}

\newcommand\Tt{\rule{0pt}{2.6ex}}
\newcommand\Bb{\rule[-1.2ex]{0pt}{0pt}}

\usepackage{amsmath} 

\usepackage{graphicx} 
\usepackage{dcolumn}  
\usepackage{bm}       

\usepackage{ifpdf}
\ifpdf
\usepackage[pdftex]{hyperref}
\else
\usepackage[hypertex]{hyperref}
\fi

\hypersetup{
 pdftitle={},%
 pdfauthor={},%
 pdfsubject={},%
 pdfkeywords={},%
 pdfstartview={},%
 bookmarksopen=true, breaklinks=true, debug=true,
 colorlinks=true, linkcolor=blue, citecolor=red, urlcolor=red,
 hyperfigures=true
}

\begin{document}

\preprint{CLNS 10/2071}
\preprint{CLEO 10-08}

\title{\boldmath Measurements of branching fractions 
for \\
electromagnetic transitions
involving the $\chi_{bJ}(1P)$ states}

\author{M.~Kornicer}
\author{R.~E.~Mitchell}
\author{C.~M.~Tarbert}
\affiliation{Indiana University, Bloomington, Indiana 47405, USA }
\author{D.~Besson}
\affiliation{University of Kansas, Lawrence, Kansas 66045, USA}
\author{T.~K.~Pedlar}
\affiliation{Luther College, Decorah, Iowa 52101, USA}
\author{D.~Cronin-Hennessy}
\author{J.~Hietala}
\author{P.~Zweber}
\affiliation{University of Minnesota, Minneapolis, Minnesota 55455, USA}
\author{S.~Dobbs}
\author{Z.~Metreveli}
\author{K.~K.~Seth}
\author{A.~Tomaradze}
\author{T.~Xiao}
\affiliation{Northwestern University, Evanston, Illinois 60208, USA}
\author{S.~Brisbane}
\author{L.~Martin}
\author{A.~Powell}
\author{P.~Spradlin}
\author{G.~Wilkinson}
\affiliation{University of Oxford, Oxford OX1 3RH, UK}
\author{H.~Mendez}
\affiliation{University of Puerto Rico, Mayaguez, Puerto Rico 00681}
\author{J.~Y.~Ge}
\author{D.~H.~Miller}
\author{I.~P.~J.~Shipsey}
\author{B.~Xin}
\affiliation{Purdue University, West Lafayette, Indiana 47907, USA}
\author{G.~S.~Adams}
\author{D.~Hu}
\author{B.~Moziak}
\author{J.~Napolitano}
\affiliation{Rensselaer Polytechnic Institute, Troy, New York 12180, USA}
\author{K.~M.~Ecklund}
\affiliation{Rice University, Houston, Texas 77005, USA}
\author{J.~Insler}
\author{H.~Muramatsu}
\author{C.~S.~Park}
\author{L.~J.~Pearson}
\author{E.~H.~Thorndike}
\author{F.~Yang}
\affiliation{University of Rochester, Rochester, New York 14627, USA}
\author{S.~Ricciardi}
\affiliation{STFC Rutherford Appleton Laboratory, Chilton, Didcot, Oxfordshire, OX11 0QX, UK}
\author{C.~Thomas}
\affiliation{University of Oxford, Oxford OX1 3RH, UK}
\affiliation{STFC Rutherford Appleton Laboratory, Chilton, Didcot, Oxfordshire, OX11 0QX, UK}
\author{M.~Artuso}
\author{S.~Blusk}
\author{R.~Mountain}
\author{T.~Skwarnicki}
\author{S.~Stone}
\author{J.~C.~Wang}
\author{L.~M.~Zhang}
\affiliation{Syracuse University, Syracuse, New York 13244, USA}
\author{G.~Bonvicini}
\author{D.~Cinabro}
\author{A.~Lincoln}
\author{M.~J.~Smith}
\author{P.~Zhou}
\author{J.~Zhu}
\affiliation{Wayne State University, Detroit, Michigan 48202, USA}
\author{P.~Naik}
\author{J.~Rademacker}
\affiliation{University of Bristol, Bristol BS8 1TL, UK}
\author{D.~M.~Asner}
\altaffiliation[Now at: ]{Pacific Northwest National Laboratory, Richland, WA 99352}
\author{K.~W.~Edwards}
\author{K.~Randrianarivony}
\author{G.~Tatishvili}
\altaffiliation[Now at: ]{Pacific Northwest National Laboratory, Richland, WA 99352}
\affiliation{Carleton University, Ottawa, Ontario, Canada K1S 5B6}
\author{R.~A.~Briere}
\author{H.~Vogel}
\affiliation{Carnegie Mellon University, Pittsburgh, Pennsylvania 15213, USA}
\author{P.~U.~E.~Onyisi}
\author{J.~L.~Rosner}
\affiliation{University of Chicago, Chicago, Illinois 60637, USA}
\author{J.~P.~Alexander}
\author{D.~G.~Cassel}
\author{S.~Das}
\author{R.~Ehrlich}
\author{L.~Fields}
\author{L.~Gibbons}
\author{S.~W.~Gray}
\author{D.~L.~Hartill}
\author{B.~K.~Heltsley}
\author{D.~L.~Kreinick}
\author{V.~E.~Kuznetsov}
\author{J.~R.~Patterson}
\author{D.~Peterson}
\author{D.~Riley}
\author{A.~Ryd}
\author{A.~J.~Sadoff}
\author{X.~Shi}
\author{W.~M.~Sun}
\affiliation{Cornell University, Ithaca, New York 14853, USA}
\author{J.~Yelton}
\affiliation{University of Florida, Gainesville, Florida 32611, USA}
\author{P.~Rubin}
\affiliation{George Mason University, Fairfax, Virginia 22030, USA}
\author{N.~Lowrey}
\author{S.~Mehrabyan}
\author{M.~Selen}
\author{J.~Wiss}
\affiliation{University of Illinois, Urbana-Champaign, Illinois 61801, USA}
\author{J.~Libby}
\affiliation{Indian Institute of Technology Madras, Chennai, Tamil Nadu 600036, India}
\collaboration{CLEO Collaboration}
\noaffiliation

\date{December 2, 2010}

\begin{abstract} 
Using $(9.32,5.88)$ million $\Upsilon(2S,3S)$ decays taken with the 
CLEO III detector, 
we obtain five product branching fractions for the exclusive processes
$\Upsilon(2S)\to \gamma \chi_{b0,1,2}(1P) \to \gamma \gamma \Upsilon(1S)$ and
$\Upsilon(3S)\to \gamma \chi_{b1,2}(1P) \to \gamma \gamma \Upsilon(1S)$.
We observe the transition 
$\chi_{b0}(1P)\to\gamma\Upsilon(1S)$ for the first time.
Using the known branching fractions for 
$\brf[\Upsilon(2S)\to\gamma\chi_{bJ}(1P)]$,
we  extract values for
$\brf[\chi_{bJ}(1P)\to\gamma\Upsilon(1S)]$ for $J=0$, $1$, $2$.
In turn, these values can be used to unfold the $\Upsilon(3S)$
product branching fractions to obtain values for
$\brf[\Upsilon(3S)\to\gamma\chi_{b1,2}(1P)]$ for the first time
individually.
Comparison of these with each other and with the branching fraction
$\brf[\Upsilon(3S)\to\gamma\chi_{b0}]$ previously measured by CLEO
provides tests of relativistic corrections to electric dipole matrix elements.

\end{abstract}

\maketitle

\section{Introduction}

The bottomonium ($b \bar b$) resonances display a rich pattern of
electromagnetic transitions, including electric dipole (E1) transitions
between $S$-wave ($\Upsilon(nS)$) and $P$-wave ($\chi_b(nP)$) states
\cite{Eichten}. Branching fractions for these transitions involving the lowest
$\chi_b(1P)$ states are summarized in Table \ref{tab:E1trans} \cite{PDG}.
Notable in Table \ref{tab:E1trans} is the suppression of the transitions
from the $\Upsilon(3S)$ to the $\chi_{bJ}(1P)$ states.  The electric dipole
matrix element for the $\Upsilon(3S)\to\gamma\chi_{bJ}(1P)$ transition,
$\langle 1P | r | 3S \rangle$, is very small 
(see Ref.\ \cite{Grant:1995hf} for a discussion), and thus is quite
sensitive to assumed shapes of wave functions due to various
relativistic corrections.  In the non-relativistic limit it should
be independent of $J$, with the corresponding decay rates given by
\[
\Gamma[\Upsilon(3S) \to \gamma \chi_{bJ}(1P)] =  
  \frac{4}{3^5}~\alpha~(2J+1)~E_\gamma^3~|\langle 1P | r | 3S \rangle|^2~.
\]
Thus, one would expect the rates for $J=0:1:2$ to be governed by the
term $E_\gamma^3 \times (2J+1)$, or to be in the ratio 1:2.4:3.6.  Various
treatments of relativistic corrections to these decay rates \cite{thmodels}
imply ratios differing considerably from these values and from one another,
providing an opportunity to distinguish between them.

\begin{table}[h]
\caption{
Previous data~\cite{PDG} for
branching fractions for electric dipole transitions
in the bottomonium ($b \bar b$) system involving the lowest P-wave spin-triplet
states $\chi_b(1P)$.  Note the absence of measurements 
for
$\Upsilon(3S) \to
\gamma \chi_{b1,2}(1P)$ and $\chi_{b0}(1P) \to \gamma \Upsilon(1S)$.
\label{tab:E1trans}}
\begin{center}
\begin{tabular}{c c c} \hline \hline
Transition \Tt \Bb & $E_\gamma$ (MeV) & $\brf$ (\%) \\ 
\hline
$\Upsilon(3S) \to \gamma \chi_{b0}(1P)$ \Tt & 483.9 & $0.30 \pm 0.11$ \\
$\Upsilon(3S) \to \gamma \chi_{b1}(1P)$    & 452.1 & $< 0.17$ \\
$\Upsilon(3S) \to \gamma \chi_{b2}(1P)$ \Bb & 433.5 & $<1.9$ \\ 
\hline
$\Upsilon(2S) \to \gamma \chi_{b0}(1P)$ \Tt & 162.5 & $3.8 \pm 0.4$ \\
$\Upsilon(2S) \to \gamma \chi_{b1}(1P)$    & 129.6 & $6.9 \pm 0.4$ \\
$\Upsilon(2S) \to \gamma \chi_{b2}(1P)$ \Bb & 110.4 & $7.15 \pm 0.35$ \\ 
\hline
$\chi_{b0}(1P) \to \gamma \Upsilon(1S)$ \Tt & 391.1 & $< 6$ \\
$\chi_{b1}(1P) \to \gamma \Upsilon(1S)$    & 423.0 & $35 \pm 8$ \\
$\chi_{b2}(1P) \to \gamma \Upsilon(1S)$ \Bb & 441.6 & $22 \pm 4$ \\ 
\hline \hline
\end{tabular}
\end{center}
\end{table}

As seen in Table \ref{tab:E1trans}, the four photon energies for the cascade
transitions $\Upsilon(3S) \to \gamma \chi_{b1,2}(1P) \to \gamma \gamma
\Upsilon(1S)$ are all in the range 423--452 MeV in the rest frames of the
decaying particle, making it difficult to extract the product branching
fractions for individual values of $J$ \cite{Heintz:1992,Skwarnicki:2002bp}.  
Thus, until now, only the sum over $J$ values (assumed here to be $J=1,2$)
has been measured.  For the summed product branching ratio,
\[
\begin{split}
\brf_{\rm{sum}}=
\sum_{J=1,2} \brf[\Upsilon(3S) \to \gamma \chi_{bJ}(1P)]\times 
\brf[\chi_{bJ}(1P) \to \gamma \Upsilon(1S)],
\end{split}
\]
Ref.\ \cite{Heintz:1992} obtains $\brf_{\rm{sum}}=(1.2^{+0.4}_{-0.3} \pm 0.09) \times 10^{-3}$,
while Ref.\ \cite{Skwarnicki:2002bp} obtains $\brf_{\rm{sum}}=(2.14 \pm 0.22 \pm 0.21) \times
10^{-3}$.

In this article, we obtain separate product branching fractions for $J=1$ and
$J=2$.  Notice, however, that unfolding $\brf[\Upsilon(3S)\to\gamma\chi_{b1,2}]$
from the product branching fractions requires knowledge
of the $\chi_{b1,2}(1P)\to\gamma\Upsilon(1S)$ rates, which, as seen in 
Table~\ref{tab:E1trans}, have relative errors exceeding $20\%$.
Therefore we also measure similar product branching fractions for
$\Upsilon(2S)$ transitions, 
$\brf[\Upsilon(2S)\to\gamma\chi_{bJ}(1P)]\times$
$\brf[\chi_{bJ}(1P)\to\gamma\Upsilon(1S)]$,
from which the E1 rates for $\chi_{bJ}(1P)$ decays can be 
extracted by using the known $\Upsilon(2S)$ branching fractions
in Table~\ref{tab:E1trans}. Thus, with product branching fractions
for two-photon cascade transitions from $\Upsilon(3S)$ and
$\Upsilon(2S)$ to $\Upsilon(1S)$ through $\chi_{bJ}(1P)$, we can make 
the first determinations of E1 branching fractions
from $\Upsilon(3S)$ through $\chi_{bJ}(1P)$ for $J=1$ and $J=2$ and
for $\chi_{b0}(1P)$ E1 transitions to $\Upsilon(1S)$, as well as
improved values for $\chi_{b1,2}(1P)\to\gamma\Upsilon(1S)$.
The branching fractions 
$\brf[\Upsilon(3S) \to \gamma \chi_{b1,2}]$ may be
compared with those
for $J=0$ previously measured by CLEO~\cite{chib},
$\brf[\Upsilon(3S) \to \gamma \chi_{b0}(1P)] = (0.30 \pm 0.04 \pm 0.10)\%$,
providing tests of relativistic corrections to electric dipole matrix elements
such as those involved in the predictions of Refs.\ \cite{thmodels}.  

We discuss the data samples in 
Sec.\ \ref{sec:datasample}.
We describe our analysis method in Sec.\ \ref{sec:method}, and 
our fits to the transitions 
$\Upsilon(2S) \to \gamma \chi_{bJ}(1P) \to \gamma \gamma \Upsilon(1S)$ and
$\Upsilon(3S) \to \gamma \chi_{bJ}(1P) \to \gamma \gamma \Upsilon(1S)$
in Secs.\ \ref{sec:2S} and \ref{sec:3S}, respectively.
We summarize our results and
compare theoretical predictions against 
the measured branching fractions 
$\brf[\Upsilon(3S)\to\gamma\chi_{bJ}(1P)]$ in Sec.\ \ref{sec:result}
and against $\brf[\chi_{bJ}\to\gamma\Upsilon(1S)]$ in Appendix~\ref{sec:appb}.

\section{Data sample}\label{sec:datasample}

Our event selection criteria (and this entire analysis procedure) closely
follow the analysis of $\Upsilon(2S) \to \eta \Upsilon(1S)$~\cite{etaart}.
The data used in this
analysis were collected in $e^+ e^-$ collisions at the
Cornell Electron Storage Ring (CESR), at center-of-mass energies at and 
$\sim 25$~MeV below the $\Upsilon(1S)$, $\Upsilon(2S)$, and 
$\Upsilon(3S)$ resonances.  Integrated luminosities 
as well as the estimated numbers of these narrow resonance decays
are shown in Table~\ref{tab:lumi}.
Events were
recorded in the CLEO III detector, equipped with an electromagnetic calorimeter
consisting of 7784 thallium-doped cesium iodide (CsI) crystals and covering
93\% of solid angle, initially installed in the CLEO II \cite{CLEO2} detector
configuration.  The energy resolution of the crystal calorimeter is 5\% (2.2\%)
for 0.1 GeV (1 GeV) photons.  The CLEO III tracking system \cite{CLEO3trk}
consists of a silicon strip detector and a large drift chamber, achieving a
charged particle momentum resolution of 0.35\% (1\%) at 1 GeV/$c$ (5 GeV/$c$)
in a 1.5 T axial magnetic field.

\begin{table}[h]
\caption{Integrated luminosities of data sets used in this analysis,
in units of pb$^{-1}$.
``ON" corresponds to data sets taken in the vicinity of the
nominal masses~\cite{PDG} of the corresponding
narrow resonances, while ``OFF" represents data taken $\sim 25$~MeV below the
respective resonance masses.
The estimated numbers of narrow resonance decays 
in each data set~\cite{upsart}
are listed at the bottom row.
\label{tab:lumi}}
\def\1#1#2#3{\multicolumn{#1}{#2}{#3}}
\begin{center}
\begin{tabular*}{0.48\textwidth}{@{\extracolsep{\fill}}c c c c} \hline \hline
  & $\Upsilon(1S)$ \Tt \Bb & $\Upsilon(2S)$ & $\Upsilon(3S)$\\
\hline
ON \Tt  & $1056$ & $1305$ & $1387$ \\
OFF \Bb & $190$ & $438$ & $158$ \\
\hline
\#$\Upsilon(nS)$ ($10^6$) \Tt \Bb 
& $20.81\pm0.37$&$9.32\pm0.14$&
$5.88\pm0.10$\\
\hline \hline
\end{tabular*}
\end{center}
\end{table}

\section{Analysis method}\label{sec:method}

One can choose a set of $\ell^+ \ell^- \gamma \gamma$ events
($\ell^{\pm}=e^{\pm}$ or $\mu^{\pm}$)
 in which the two
photon energies sum up to a range consistent with the transition
$\Upsilon(2S,3S) \to \gamma \chi_{bJ}(1P) \to \gamma \gamma \Upsilon(1S)$.
We label the energy of the lower-energy photon
$\egl$, and that of the higher-energy photon $\egh$.
In this section, we describe how we select our lepton candidates
from $\Upsilon(1S)\to\ell^+\ell^-$, how we apply kinematic constraints, and
what our main backgrounds are.

\subsection{Selection of leptons and photons}\label{sec:evtselect}

In order to identify leptonic decays of $\Upsilon(1S)$,
we first select the two highest-momentum tracks in an event.  
We call a track an
electron candidate if $E/p > 0.75$ or a muon candidate if $E/p < 0.20$, where
$E$ is the energy observed in the calorimeter 
shower associated with it and $p$ is its momentum measured
in the tracking system.  Each track must satisfy
$|\cos \theta| < 0.83$, where $\theta$ is the angle with respect to the 
positron beam direction, to ensure reliable triggering and optimal 
performances of the tracking system and calorimeter for lepton identification.
These tracks must originate within 5 cm (5 mm) along the beam direction
(in the $r$--$\phi$ plane) of the the interaction point (IP).
Both tracks must be of the same lepton type and be of opposite charge.

Electron candidate tracks are dealt with somewhat differently,
as they may 
radiate energy via bremsstrahlung, and also contain significant contamination 
from radiative Bhabha scattering.  To recover bremsstrahlung, we add to each
lepton candidate's four-momentum the four-momentum 
of any photon candidates 
found to lie within a cone of 100 mrad of the lepton candidate track 
direction at the IP.
To suppress contributions from Bhabha scattering in 
$\gamma \gamma e^+e^-$ final
states, we require $e^+$ candidates to satisfy $\cos \theta_{e^+} < 0.5$, where
the final state positron makes an angle $\theta_{e^+}$ with the 
incoming positron beam direction.
This selection criterion greatly suppresses Bhabha scattering background
while keeping a large fraction of the signal.

Photon candidates must be detected either in 
the
barrel ($|\cos \theta| < 0.81$) or in 
the
endcaps ($0.85<|\cos \theta| < 0.93$) of
the calorimeter. Each
must have a lateral shower profile consistent with that of a photon, and
the shower energy must exceed 30 (50) MeV in the barrel (endcaps).  
Additionally,
such showers must not be aligned with the initial momentum of a track.

\subsection{Background composition}\label{sec:bkgsample}

As in the analysis of $\Upsilon(2S,3S)\to\pi^0/\eta\Upsilon(1S,2S)$
\cite{etaart}, the dominant (and almost sole) sources of background are
the doubly-radiative QED processes $e^+e^-\to\ell^+\ell^-\gamma\gamma$.
Such events can completely satisfy the restrictions on
kinematic fit quality (see below) when the $\ell^+\ell^-$ coincidentally
has an invariant mass near that of the $\Upsilon(1S)$.
Using our off-resonance data samples described in Table~\ref{tab:lumi}
to study these backgrounds, we find that such events produce
smooth, nearly flat, nonpeaking spectra in $\egl$ and $\egh$.
For the $\Upsilon(3S)\to\gamma\gamma\ell^+\ell^-$ analysis,
we prepare 
Monte Carlo
(MC) simulations of these processes for use
in the nominal fits (see 
Sec.\ \ref{sec:qedbkg} ).

To improve photon resolutions, we use values of
kinematically constrained four-vectors instead of
the observed ones.
We perform kinematic fitting on
events in the following manner (cf.\ Ref.\ \cite{etaart}). 
We constrain the two leptons to have the $\Upsilon(1S)$ mass
and the total four-momentum of the $\Upsilon(1S)$ candidate
and the two photons to be that of the incoming $e^+e^-$,
accounting for the non-zero crossing angle of the beams.
The dilepton vertex is constrained to the beam spot,
which itself is measured with multi-track hadronic events
for each run of approximately 1 hour duration. The photons
are assumed to originate at the beam spot.

The specific procedure employed is the following:
Obtain reduced $\chi^2$ values (i.e., values of $\chi^2$ per degree of freedom)
from the above vertex and momentum fits. 
Call them $\chi^2_{v,1C}$ and $\chi^2_{m,1C}$, respectively.
Combine the mass-fitted object with two-photon candidates and constrain
the sum of their four-momenta
(a 4C fit) to the laboratory four-momentum,
obtaining a reduced $\chi^2$ value denoted by $\chi^2_{m,4C}$.
Require reduced $\chi^2$ values in the following order:
 (1) $\chi^2_{v,1C}<10$; (2) $\chi^2_{m,1C} < 10$ 
 (3) $\chi^2_{m,4C} <5$.

Our kinematic fitting software assumes that measured distributions are
Gaussian, but there are well-understood and reasonably well-modeled
low-side tails on the energy response of the calorimeter.  Hence,
the (reduced) $\chi^2$ distributions from fits in both MC and data will
also have tails not seen in formal $\chi^2$ probability distributions. Good
agreement of the reduced $\chi^2$ distributions between data and MC is
essential. We explored this agreement in previous CLEO
analyses~\cite{CLEOcuts} and in this analysis (Sec.\ \ref{sec:kfitsyst}). 
We included the small discrepancies we found in our systematic errors.

\begin{figure}[htbp]
\begin{center}
\scalebox{1}
{\includegraphics[width=0.82\textwidth]{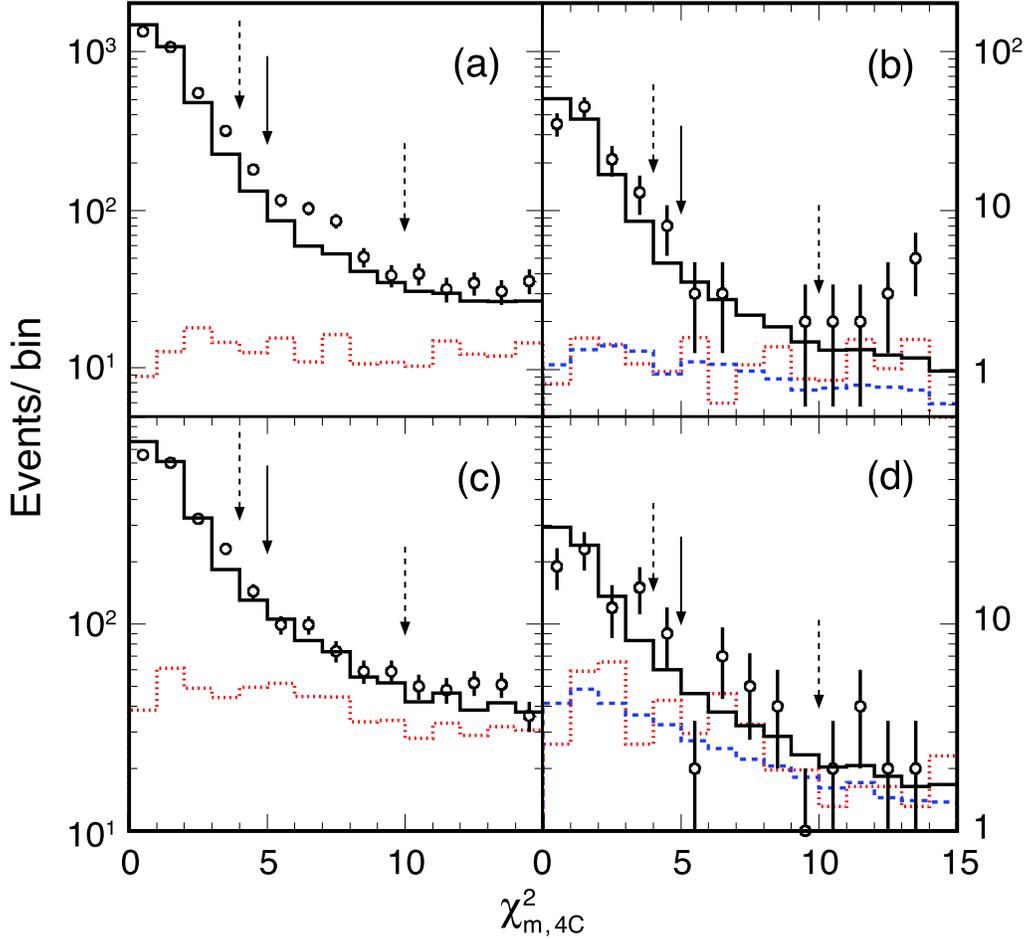}
}
\caption{Distributions of $\chi^2_{m,4C}$ for 
$\Upsilon(1S)\to \mu^+\mu^-$ 
candidates
(a,b) and 
$\Upsilon(1S)\to e^+e^-$ 
candidates (c,d)
based on events of 
$\Upsilon(2S)\to\gamma\gamma\Upsilon(1S)$ 
candidates (a,c) and
$\Upsilon(3S)\to\gamma\gamma\Upsilon(1S)$ 
candidates (b,d)
via $\chi_{bJ}(1P)$, respectively.
Open circles represent data, solid histograms
the sum of scaled signal MC and background contributions,
where the background levels are indicated by
scaled off-resonance data (dotted histogram) and,
for $\Upsilon(3S)$ decays, QED MC simulation (dashed histograms).
Solid arrows indicate 
standard
selection criteria,
and dashed arrows alternate values used for systematic
studies of dependence 
on selection criteria.
\label{fig:c24cmom}}
\end{center}
\end{figure}

Figure\ \ref{fig:c24cmom} shows distributions of one of our reduced
$\chi^2$ variables, $\chi^2_{m,4C}$, for the 
$\Upsilon(2S,3S)\to\gamma\gamma\Upsilon(1S)$ analyses.
Standard and alternate restrictions on these variables
are indicated in the figure. 
The overlaid histograms indicate the contributions
of MC signal and background, weighted by our final measured values.
Based on the scaled off-resonance data, the expected background levels 
in the signal regions correspond to background-to-signal
ratios of $\sim 1.5~(3)\%$ for the $\Upsilon(1S) \to \mu^+ \mu^-$ 
candidates and $\sim 15~(20)\%$ for the $\Upsilon(1S) \to e^+ e^-$ 
candidates in $\Upsilon(2S)$ ($\Upsilon(3S)$) decays.

\section{\boldmath Analysis of $\Upsilon(2S)\to\gamma\gamma\ell^+\ell^-$ via
$\chi_{bJ}(1P)$}\label{sec:y2s1p1s} \label{sec:2S}

\begin{figure}
\begin{center}
\scalebox{1}
{
\includegraphics[width=0.49\textwidth]{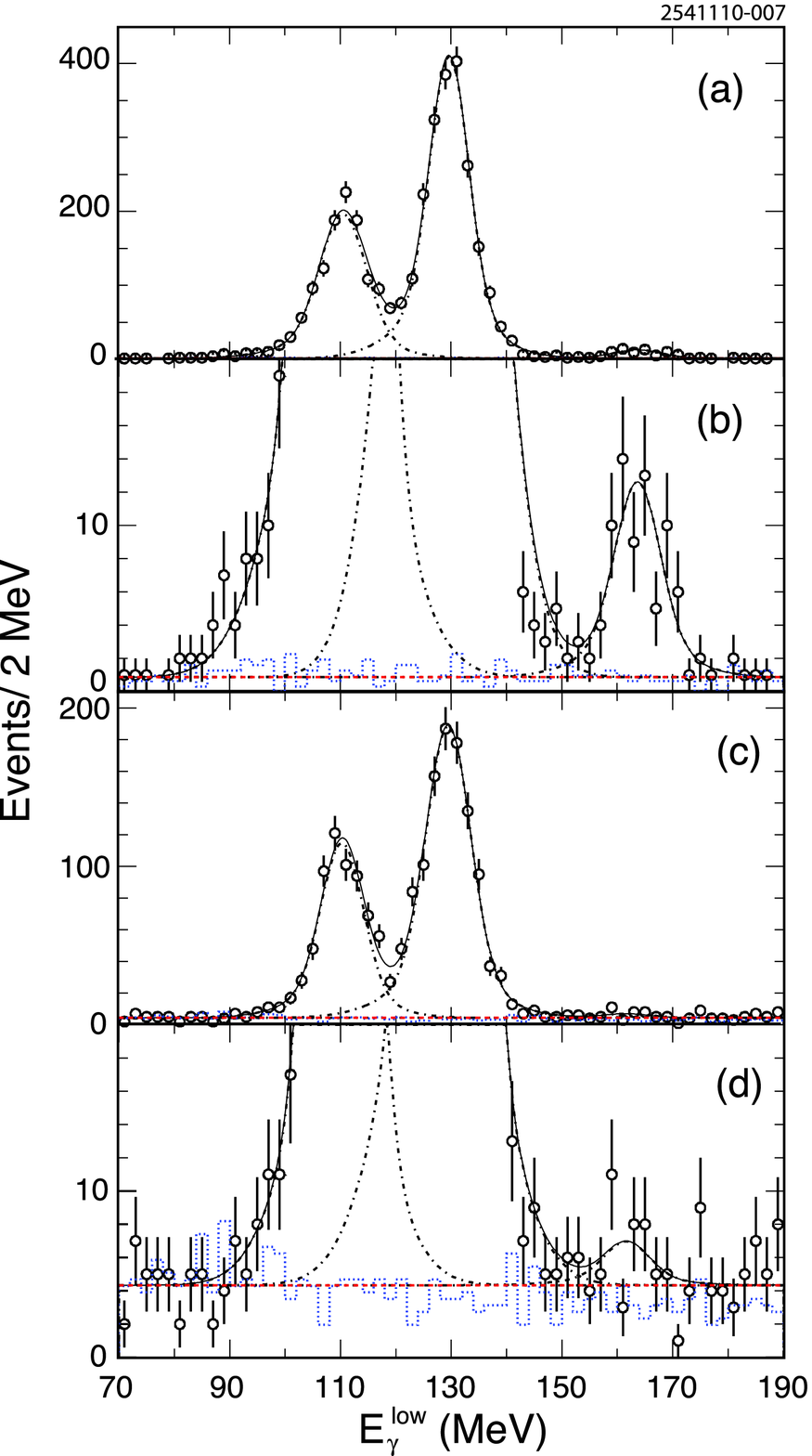}
}
\caption{Fits to data with a flat background shape represented by 
dashed histograms for $\Upsilon(1S)\to \mu^+\mu^-$ $[e^+e^-]$ candidates in the
two plots (a,b) [(c,d)].  Plots (b) and (d) are identical,
respectively, to (a) and (c) but zoomed 
in so as to highlight the $J=0$ component.
The dotted histograms represent the scaled
background shapes based on the off-resonance data. 
The dot-dashed lines show components of each of the three photon lines.
\label{fig:y2sdata}}
\end{center}
\end{figure}

Because the expected $\egl$ spectra ($110$--$160$~MeV)
have excellent separation of decays through
different $\chi_{bJ}(1P)$ spin states, we can obtain
the yields of the individual $J$ contributions with
a fit to $\egl$, without regard to $\egh$ ($390$--$440$~MeV). 
The $\egl$ spectra exhibit three clearly distinguishable peaks
with known peak energies and resolutions entirely
dominated by measurement effects, although 
detector resolution is improved upon
by the constrained fit.
Figure~\ref{fig:y2sdata} illustrates the $\egl$ distributions
from data, along with the fits described in the following subsections.

\subsection{Signal Monte Carlo samples}\label{sec:anamc2s}

We use the EVTGEN event generator~\cite{evtgen} to generate expected
signal shapes in Monte Carlo (MC) simulations. To simulate the photon angular
distributions, we assume that the radiative transitions are pure
electric dipole. We generate $100$~k events with
$\brf[\Upsilon(2S)\to\gamma\chi_{bJ}(1P)]=$ $\brf[\chi_{bJ}\to\gamma\Upsilon(1S)]
=100\%$ while $\brf[\Upsilon(1S)\to e^+e^-]=\brf[\Upsilon(1S)\to\mu^+\mu^-]=50\%$
for each of $J=0$, $1$, and $2$. We fit each of the MC distributions of
$\egl$ with 
a double Gaussian whose difference between the two means we allow to float.  
Resultant reconstruction efficiencies are shown in Table~\ref{tab:2sfitresult}.

\begin{table}[htbp]
\caption{Fitted yields,
reconstruction efficiencies ($\epsilon$),
and corresponding branching fractions are shown with statistical errors only.
Here,
$\brf1 = \brf[\Upsilon(2S)\to\gamma\chi_{bJ}(1P)]$,
$\brf2 = \brf[\chi_{bJ}(1P)\to\gamma\Upsilon(1S)]$, and
$\brf3 = \brf[\Upsilon(1S)\to\ell^+\ell^-]$.
The last column (``All $J$'') shows 
results for the sum of $J=0$, $1$, and $2$ components obtained by
subtracting fits to the backgrounds from the data,
as described in the text.
\label{tab:2sfitresult}}
\def\1#1#2#3{\multicolumn{#1}{#2}{#3}}
\begin{center}
\begin{tabular*}{0.97\textwidth}{@{\extracolsep{\fill}}c c c c c c}
\hline \hline
 \Tt \Bb & \1{1}{c}{$\ell^+\ell^-$}&\1{1}{c}{$J=0$}&\1{1}{c}{$J=1$}&\1{1}{c}{$J=2$} &
   \1{1}{c}{All $J$} \\ 
\hline
Yields \Tt & $e^+e^-$        &$16\pm9$  &$1068\pm36$&$600\pm30$  &$1684\pm41$ \\
Yields \Bb & $\mu^+\mu^-$&$71\pm10$&$2154\pm50$&$1170\pm39$&$3395\pm58$ \\
\hline
$\epsilon~(10^{-2})$ \Tt &$e^+e^-$ 
&$20.9\pm0.2$&$21.3\pm0.2$&$20.2\pm0.2$&$20.9\pm0.2$\\
$\epsilon~(10^{-2})$ \Bb &$\mu^+\mu^-$
&$38.6\pm0.3$&$39.9\pm0.3$&$37.7\pm0.3$&$39.1\pm0.2$\\
\hline
$\brf1\times\brf2\times\brf3~(10^{-4})$ \Tt &$e^+e^-$
&$0.083\pm0.044$&$5.38\pm0.18$&$3.19\pm0.16$&$8.66\pm0.21$ \\
$\brf1\times\brf2\times\brf3~(10^{-4})$ \Bb &$\mu^+\mu^-$
&$0.196\pm0.028$&$5.79\pm0.13$&$3.33\pm0.11$&$9.32\pm0.16$\\
\hline
$\brf1\times\brf2\times\brf3~(10^{-4})$ \Tt \Bb &$e^+e^-$ and $\mu^+\mu^-$
&$0.163\pm0.024$&$5.65\pm0.11$&$3.29\pm0.09$&$9.08\pm0.13$\\
\hline \hline
\end{tabular*}
\end{center}
\end{table}

\subsection{Fitting the data}

Our nominal fit procedure is to take these double Gaussian shapes based on the
signal MC samples to fit to data,
fixing the respective narrower Gaussian widths and differences between
the two Gaussian means but allowing
the larger widths ($\sigma_J$ where $J=0$, $1$, and $2$)
to float.
We use a flat background shape whose normalization is also
allowed to float.
We then perform a maximum likelihood fit.
  Figure~\ref{fig:y2sdata} shows fits to data for
$\Upsilon(1S)\to \mu^+\mu^-$ candidates 
(a,b) and
$\Upsilon(1S)\to e^+e^-$ candidates (c,d)
with flat background
shapes represented by the dashed histograms,
while the dotted histograms are based on 
scaled off-$\Upsilon(2S)$-resonance data.
In Fig.\ \ref{fig:y2sdata}(b), we zoom in to a smaller
vertical scale to emphasize the $J=0$ component,
which is clearly visible for the $\mu^+\mu^-$ candidates.
However, in Fig.\ \ref{fig:y2sdata}(d)
the $J=0$ peak is obscured by the relatively larger
background in the $e^+e^-$ candidates.
To compensate for the larger backgrounds in the distributions
for the the $e^+e^-$ candidates, we fix the ratio of the widths
of the $J=0$ peak to the width of the $J=1$ peak
($\sigma_0/\sigma_1$) to be equal to the ratio obtained 
for the signal MC samples.
(In our systematic study, we remove this restriction and observe the
deviation from our central value.)
The observed yields, along with efficiency-corrected 
products of branching fractions, are shown in Table~\ref{tab:2sfitresult}.
The last column of Table~\ref{tab:2sfitresult} (labeled as ``All $J$'')
represents measurements of these same two photon cascade events, 
but is summed over $J$ for $J=0$, $1$, and $2$. We obtain the yield simply by
subtracting the fitted background shape, with normalization determined
from the nominal fit, and then by 
summing the resultant spectrum over the signal region.
There, efficiencies are weighted by the measured
branching fractions for each spin state.
The $\chi^2$ values for the fits are 50.1 for 60 data points (minus 9
parameters), c.l.\ = 51.0\% for $\Upsilon(1S) \to e^+ e^-$, and 51.4 for
60 data points (minus 10 parameters), c.l.\ = 42.0\% for $\Upsilon(1S) \to
\mu^+ \mu^-$.

\subsection{Systematic uncertainties~\label{subsec:y2ssys}}

Systematic uncertainties are assessed as
$J$-independent contributions that impact
the yields of several modes equally, or are determined
individually. Where errors differ for $e^+e^-$ and $\mu^+\mu^-$
candidates, they are averaged with the same statistical
weights used for combining the respective product branching fractions. 

\subsubsection{Uncertainties common to both $e^+e^-$ and $\mu^+\mu^-$ 
candidates}

The uncertainties common to both $e^+e^-$ and $\mu^+\mu^-$ candidates
are relative uncertainties known already from external
sources, and which impact every yield identically.  The relative uncertainties
on the numbers of resonance decays in our data sets were estimated in Ref.\ 
\cite{upsart}.  We take the dilepton reconstruction systematic uncertainty from
Ref.\ \cite{pipiart} as $1.0\%$, since our dilepton reconstruction is
identical to that used there.  Similarly we use the result from that paper for
$\piz$-finding, which was $1.6\%$ per $\piz$.  We therefore use this as an
estimate of the uncertainty for reconstructing the two photons in the cascade,
since the photon energies we are studying in this analysis very closely
resemble the photon energies in Ref.\ \cite{pipiart},
and otherwise the processes are kinematically very similar.
Instead of $1.6\%$ for the pair, we round up to $2.0\%$ to be conservative.

\subsubsection{Signal shapes}

In our nominal fit procedure, we take a double Gaussian, fitted to the signal
MC samples, to represent the signal line shapes
but float the larger widths of the Gaussians for the fits to 
data to accommodate imperfect
simulations of detector resolutions. Here, we have tried the
following variations:

\begin{itemize}

 \item
 Constrain all $\sigma_J$ in both $\Upsilon(1S)\to \ell^+\ell^-$ candidates
while requiring the ratios $\sigma_0/\sigma_1$ and $\sigma_2/\sigma_1$
to be the same as in our signal MC samples.

  \item Do not constrain the ``$\sigma$'' in either lepton flavor.

\end{itemize}

\subsubsection{Background shape}\label{sec:y2s-bkg-syst}

The only backgrounds predicted by a MC simulation of
all $\Upsilon(2S)$ decays are from 
$\Upsilon(2S)\to\pi^0\pi^0\Upsilon(1S)$. Even so,
this background source is found to be negligible because 
it contributes only 5-10 events to the signal regions and
it has no significant structure in photon energy.
Hence, to the extent it matters at all, it will tend
to get absorbed into the background shape in the fit.
Backgrounds from $\Upsilon(2S)\to\pi^0/\eta\Upsilon(1S)$,
using the recently measured branching fractions~\cite{etaart},
are also found to be insignificant.

To represent our background, whose main compositions are either  
doubly-radiative Bhabha events or $\mu$-pairs,
we use a flat shape with floating normalization in our nominal fit.
To probe the sensitivity to the fitted yields due to this assumption, we try a
first-order polynomial.

To determine possible systematical effects due to the fixed
background normalization in the ``All $J$'' procedure,
we vary the background by $\pm 1~\sigma$ of the statistical uncertainty
observed in our nominal procedure.

\subsubsection{Kinematic fit reduced $\chi^2$ requirements}\label{sec:kfitsyst}

We consider different choices in reduced $\chi^2$
criteria for our vertex and four-momentum fits.
The four variations to our standard selection are:
$\chi^2_{m,4C} <4$ or $<10$ rather than 5,
$\chi^2_{m,1C} < 5$ rather than 10, and
$\chi^2_{v,1C} < 5$ rather than 10.
The spin of the $\chibj$ state should have almost
no influence on how these reduced $\chi^2$ distributions are simulated,
so to increase statistical stability on the resultant variations,
we take variations from  the ``All $J$'' procedure to assign this systematic
uncertainty.

\subsubsection{Lepton flavor difference}

We have also assessed possible systematic uncertainties due to the difference
between $\ee$ and $\mumu$ results by calculating the yield for the sum of $J=1$
and $J=2$ for $\ee$ and then for $\mumu$.
(As in the case of determining possible uncertainties due to requirements on
the fitted reduced $\chi^2$, the spin of the $\chibj$ state
should have no influence on whether leptons are correctly reconstructed.) We
then took half the difference between the yields obtained for each lepton
flavor, and divided by the average yield as an estimate of the relative
systematic uncertainty arising from lepton flavor differences.

\subsubsection{Additional contributing uncertainties}

Other possible systematical effects we have investigated include: variations
in fit ranges, histogram binning, 
statistical uncertainties in signal MC samples, uncertainties in the
measured $\brf[\Upsilon(2S)\to\gamma\chi_{bJ}(1P)]\times$
$\brf[\chi_{bJ}(1P)\to\gamma\Upsilon(1S)]$ to weight efficiencies
for the ``All $J$'' case,
as well as cuts on $E/p$ to identify
lepton species.

Table~\ref{tab:syst2sll} shows sources of systematic uncertainties we have
considered for this analysis. 

\begin{table}[htbp]
\caption{Fractional uncertainties (in $\%$) 
on the combined-dilepton product branching
fractions, $\brf[\Upsilon(2S)\to\gamma\chi_{bJ}(1P)]\times$
$\brf[\chi_{bJ}(1P)\to\gamma\Upsilon(1S)]\times$
$\brf[\Upsilon(1S)\to \ell^+\ell^-]$,
due to variations of the listed
selection criteria and fit procedures.
The last column (``All $J$'') shows 
fit results for the sum of the three two-photon
cascades via $\chi_{bJ}(1P)$ states for $J=0$, $1$, and $2$. For this last
column, the entry for ``MC simulation'' includes not only statistical errors on
reconstruction efficiencies, but also the effect of total uncertainties of the
measured branching fractions on the weighted efficiency.
}
\label{tab:syst2sll}
\begin{center}
\scalebox{1.0}
{
\def\1#1#2#3{\multicolumn{#1}{#2}{#3}}
\begin{tabular*}{0.65\textwidth}{@{\extracolsep{\fill}}c c  c  c  c} \hline \hline
 Contribution    \Tt \Bb      & $J=0$  & $J=1$  & $J=2$ & All $J$\\
\hline
 $N_{\Upsilon(2S)}$  \Tt     & \1{4}{c}{$1.5$} \\
 Track-finding          & \1{4}{c}{$1.0$} \\
 Photon-finding          & \1{4}{c}{$2.0$} \\
 Reduced $\chi^2$ requirement & \1{4}{c}{$1.4$} \\
 Lepton identification   & \1{4}{c}{$0.4$} \\
 Lepton flavor difference \Bb & \1{4}{c}{$3.6$} \\
\hline
 Fit range         \Tt   & $7.1$  & $0.2$  & $0.7$ & $0.4$ \\
 Signal shape         & $1.4$  & $0.5$  & $0.6$ & --- \\
 QED bkg shape        & $1.9$  & $0.01$ & $0.2$ & $0.9$ \\
 Bin width            & $1.6$  & $0.04$ & $0.01$ & $0.01$ \\
 MC simulation   \Bb   & $0.8$  & $0.8$  & $0.8$  & $0.6$ \\
\hline
 Total           \Tt \Bb     & $9.0$ & $4.8$ & $4.9$ & $4.9$\\
\hline \hline
\end{tabular*}
}
\end{center}
\end{table}

\subsection{\boldmath Results on analysis of
$\Upsilon(2S)\to\gamma\gamma\ell^+\ell^-$}\label{sec:2sresults}

Table~\ref{tab:result2sll} shows our final results
for our $\Upsilon(2S)$ analysis
as well as those from other experiments.

\begin{table}[htbp]
\caption{Final results of this analysis. Here,
$\brf1 = \brf[\Upsilon(2S)\to\gamma\chi_{bJ}(1P)]$,
$\brf2 = \brf[\chi_{bJ}(1P)\to\gamma\Upsilon(1S)]$, and
$\brf3 = \brf[\Upsilon(1S)\to\ell^+\ell^-]$.
We use $\brf3 = (2.48\pm0.05)\%$~\cite{PDG} and
$\brf1$ values from Table~\ref{tab:E1trans}
to extract $\brf1\times\brf2$ as well
as $\brf2$.
The last column (``All $J$'') shows fit
results for the sum over $J=0$, $1$, and $2$.
Again, the first errors are statistical, the second errors are systematic, and
the third errors (when applicable) are uncertainties due to 
uncertainties in $\brf1$ and/or $\brf3$.
In the bottom half of the table
we also show results from other experiments for a comparison.
}
\label{tab:result2sll}
\begin{center}
\scalebox{0.78}
{
\def\1#1#2#3{\multicolumn{#1}{#2}{#3}}
\begin{tabular*}{1.25\textwidth}{@{\extracolsep{\fill}}c c c c c} \hline \hline
 \Tt \Bb   &$J=0$       &$J=1$   & $J=2$ & All $J$ \\
\hline
 $\brf1\times\brf2\times\brf3~(10^{-4})$ \Tt
    &$0.163\pm0.024\pm0.015$&$5.65\pm0.11\pm0.27$&$3.29\pm0.09\pm0.16$&
                             $9.08\pm0.13\pm0.44$\\
 $\brf1\times\brf2~(10^{-3})$
    &$0.659\pm0.096\pm0.059\pm0.013$
    &$22.8\pm0.4\pm1.1\pm0.5$
    &$13.3\pm0.4\pm0.6\pm0.3$
    &$36.7\pm0.6\pm1.8\pm0.7$ \\
 $\brf2~(10^{-2})$ \Bb
    &$1.73\pm0.25\pm0.16\pm0.19$
    &$33.0\pm0.6\pm1.6\pm2.0$
    &$18.5\pm0.5\pm0.9\pm1.0$
    & --- \\
\hline
\multicolumn{5}{c}{Values of $\brf2~(10^{-2})$ from other experiments \Tt} \\
 PDG average~\cite{PDG}  &$<6$ at $90\%$ CL  & $35\pm8$ & $22\pm4$ & --- \\
 Crystal Ball~\cite{cbal}&$<6$ at $90\%$ CL &$32\pm6\pm7$ &$27\pm6\pm6$& --- \\
 CUSB \Bb &$<11$ at $90\%$ CL~\cite{cusb0} 
            &$47\pm18$~\cite{cusbj}&$20\pm5$~\cite{cusbj}& --- \\
\hline \hline
\end{tabular*}
}
\end{center}
\end{table}

\section{\boldmath Analysis of $\Upsilon(3S)\to\gamma\gamma\ell^+\ell^-$ via
$\chi_{bJ}(1P)$}\label{sec:3S}

In the case of the three transitions $\upsiii\goesto\gamma\chibj\goesto\gamgam
\upsi$, the higher and lower energy photons have similar energies. Furthermore,
among these transitions in some cases the higher energy photon is emitted from
the $\upsiii$, and in some cases the higher energy photon is emitted from the
boosted $\chibj$.  Therefore,
instead of fitting just one distribution, as we do for the $\upsii$ analysis, 
we maximize our use of information and fit the two-dimensional
histogram of $\egl$ vs.\ $\egh$, where $\egl$ $<$ $\egh$ are kinematically
constrained $E_\gamma$.  Our fit will utilize
the 2D histograms from our signal Monte Carlo for each of the $J=0,1,$ and $2$
samples, as well as MC samples of doubly-radiative 
Bhabha events and $\mu$-pairs.

\subsection{Monte Carlo samples}\label{sec:anamc}

\subsubsection{Signal MC for $\upsiii\goesto\gamma\chibj$}

Approximately 100k events for each spin were generated for each of the
$\chibj$ subsamples, where $\upsiii\goesto\gamma\chibj$ at 100$\%$,
$\chibj\goesto\gamma\upsi$ at 100$\%$ and with $\upsi$ decaying half the time
to each of $\mumu$ and $\ee$.  These samples 
were used to generate the three relevant 2D histograms of
$\egl$ vs.\ $\egh$ for each spins.
In projections of these
histograms on the $\egh$ and $\egl$ axes, for $J=0$ and $J=1$, the higher-%
energy photon has a sharp distribution, while the lower-energy photon is
Doppler-broadened (see Table~\ref{tab:E1trans}).
For $J=2$, the situation is reversed to a large extent. 

\begin{figure}[hbtp]
\begin{center}
\includegraphics[width=1\textwidth]{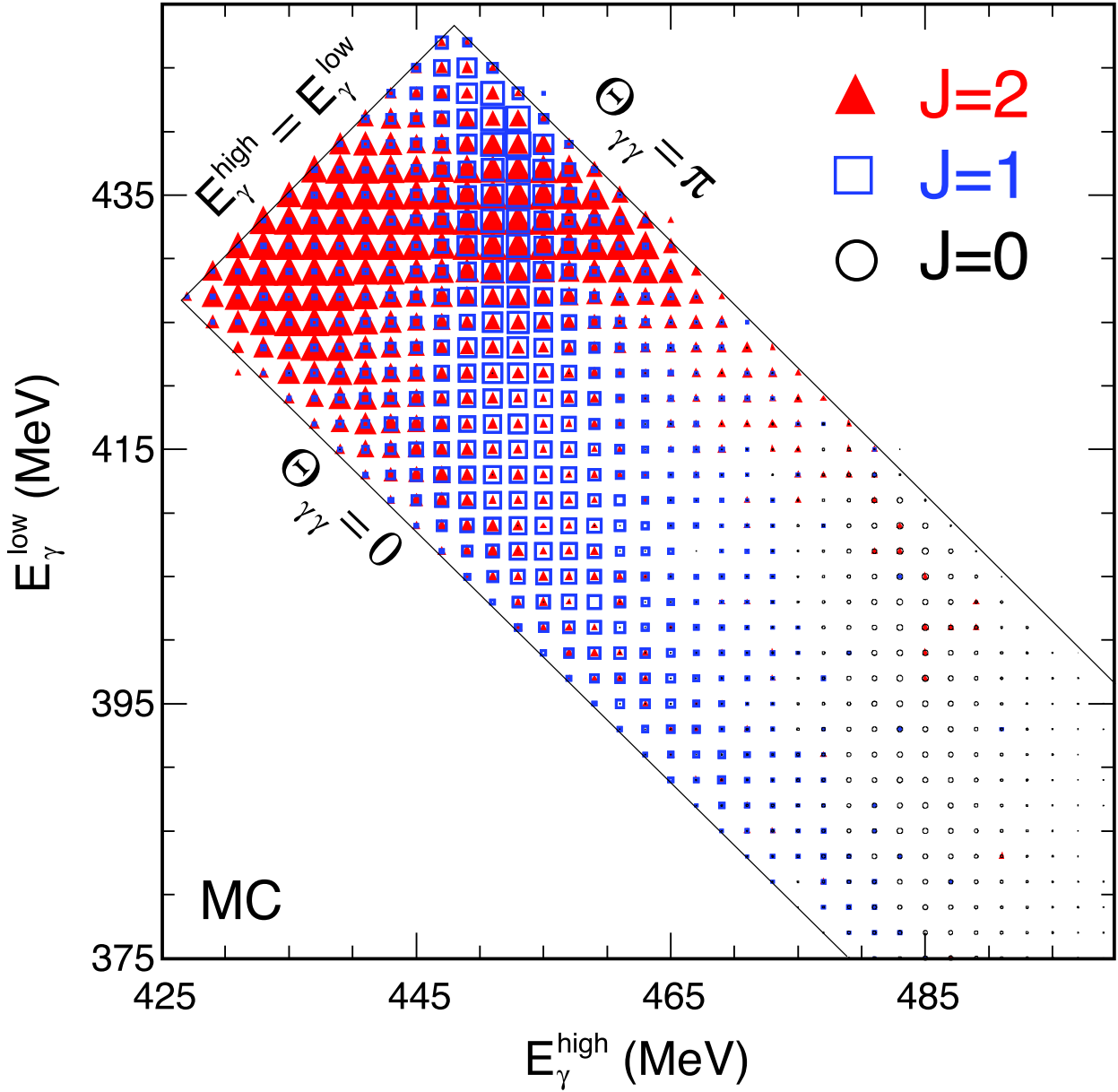}
\caption{Distributions of $\egl$ vs.\ $\egh$
based on $\Upsilon(3S)$ signal MC samples for $J=0$ (open circles), $J=1$ (open
rectangles), and $J=2$ (closed triangles) using $\mumu$ selection.
The diagonal band, edged by lines labeled as 
$\Theta_{\gamma\gamma}=\pi$
(or $0$), where $\Theta_{\gamma\gamma}$ is an opening angle between the two
emitted photons, is generated due to our kinematic constraints.  The three
samples are normalized to our measured production rates in 
this work while the size of a symbol in a bin
is proportional to the number of events for the corresponding $J$ in that bin.
\label{mc2dsigfig}}
\end{center}
\end{figure}

Figure~\ref{mc2dsigfig}
shows the $\egl$ vs.\ $\egh$ two-dimensional
histogram for $\Upsilon(3S)$ signal MC 
samples, weighted according to our final
measured branching fractions. Notice that
the events are restricted to the diagonal
band by two simple kinematic facts: first,
that $\egl<\egh$, and second, that the invariant
mass recoiling against the two photons
has been constrained to that of the 
$\Upsilon(1S)$. 
The lower boundary line corresponds to the case in which
the photon three-momenta have zero opening angle
in the initial $e^+e^-$ rest frame  (recoiling against the dilepton pair
with maximum possible combined momentum), and the upper
boundary line to the case in which the photon momenta directly oppose
one another (with the softer one traveling along the same direction
as the $\Upsilon(1S)$ candidate, and the more
energetic one in the opposite direction).
Note that the boundaries are purely kinematic in
nature due to the $\Upsilon(1S)$ mass constraint,
and hold for background as well as signal events.
In Fig\ \ref{mc2dsigfig}, these limits are labeled
$\Theta_{\gamma\gamma}=0$ and $\pi$, respectively. 

The population within the band is directly related to the
cosine of the angle $\theta_{\gamma \gamma}$ between the two photons in the
rest frame of the $\chi_{bJ}(1P)$.  
The distributions $W_J(\cos \theta_{\gamma \gamma})$ for $J=1,2$, normalized
so that their integral over $\cos \theta_{\gamma \gamma}$ is 1, are
\bea
W_1(\cos \theta_{\gamma \gamma}) & = & \frac{15}{32}(1 + \frac{1}{5}
\cos^2 \theta_{\gamma \gamma})~,\\
W_2(\cos \theta_{\gamma \gamma}) & = & \frac{1}{160}(73 + 21
\cos^2 \theta_{\gamma \gamma})~,
\eea
implying a slight enhancement at each end of the $\cos \theta_{\gamma \gamma}$
range and hence at maximum and minimum Doppler-broadened photon energy.
This feature is present in all of our signal MC samples that take
account of the photon angular distributions properly.
The respective reconstruction efficiencies are shown in
Table~\ref{tab:3sfitresult}.

\begin{table}[hbtp]
\caption{Fitted yields,
reconstruction efficiencies ($\epsilon$),
and corresponding branching fractions are shown. Here,
$\brf1 = \brf[\Upsilon(3S)\to\gamma\chi_{bJ}(1P)]$,
$\brf2 = \brf[\chi_{bJ}(1P)\to\gamma\Upsilon(1S)]$, and
$\brf3 = \brf[\Upsilon(1S)\to\ell^+\ell^-]$.
The last column (``$J=1$ and $2$'') shows results of sum of the two two-photon
cascades via $\chi_{bJ}(1P)$ states for $1$ and $2$
obtained by subtracting fits to the backgrounds from the data, as
described in the text.
\label{tab:3sfitresult}}
\def\1#1#2#3{\multicolumn{#1}{#2}{#3}}
\begin{center}
\begin{tabular*}{0.97\textwidth}{@{\extracolsep{\fill}}c c  c   c  c } 
\hline \hline
 \Tt \Bb &$\ell^+\ell^-$ &\1{1}{c}{$J=1$}&\1{1}{c}{$J=2$}&\1{1}{c}{$J=1$ and $2$}\\ 
\hline
Yields \Tt & $e^+e^-$    &$12\pm6$&$48\pm9$&$61\pm8$\\
Yields \Bb & $\mu^+\mu^-$&$38\pm9$&$78\pm11$&$117\pm11$ \\
\hline
$\epsilon~(10^{-2})$ \Tt &$e^+e^-$
             &$22.3\pm0.2$&$21.3\pm0.2$&$21.6\pm0.2$\\
$\epsilon~(10^{-2})$ \Bb &$\mu^+\mu^-$
             &$41.1\pm0.2$&$38.8\pm0.2$&$39.5\pm0.2$\\
\hline
$\brf1\times\brf2\times\brf3~(10^{-5})$ \Tt &$e^+e^-$
             &$0.91\pm0.49$&$3.88\pm0.70$&$4.79\pm0.61$\\
$\brf1\times\brf2\times\brf3~(10^{-5})$ \Bb &$\mu^+\mu^-$
             &$1.58\pm0.38$&$3.40\pm0.49$&$5.06\pm0.47$\\
\hline
$\brf1\times\brf2\times\brf3~(10^{-5})$ \Tt \Bb &$e^+e^-$ and $\mu^+\mu^-$
             &$1.33\pm0.30$&$3.56\pm0.40$&$4.96\pm0.37$\\
\hline \hline
\end{tabular*}
\end{center}
\end{table}

\subsubsection{Background from QED processes}\label{sec:qedbkg}

Only the
processes  $\ee\goesto\gamgam(\ee,\mumu)$ (and to a much lesser extent,
$\upsiii\goesto\pizpiz\upsi$)
can be reasonably expected to contribute much background.  To represent
the doubly-radiative QED events, we prepare 
MC samples generated via the Babayaga event generator~\cite{babayaga} with
sizes of roughly 100 (200) times larger luminosity
than what the data have for
doubly-radiative Bhabha ($\mu$-pair) events, respectively.
The very large number of QED events are then reduced with a very loose 
selection at the generator level to have at least two photons
and a dilepton invariant mass near that of the $\Upsilon(1S)$.
Only those chosen events are passed along to the next stage of   
processing, the CLEO detector simulation.

\begin{figure}[hbtp]
\begin{center}
\includegraphics[width=0.65\textwidth]{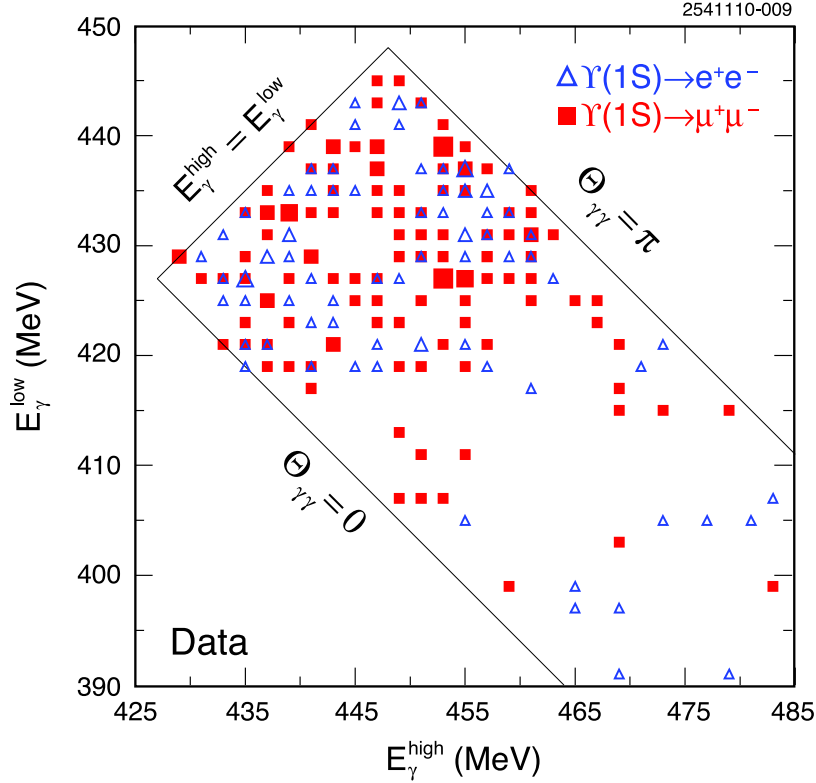}
\end{center}
\caption{Distributions of $\egl$ vs.\ $\egh$ based on the
on-$\Upsilon(3S)$ data.
Open triangles (closed rectangles) represent data points based on
$\ee$ ($\mumu$) selection.
The size of each symbol is proportional to the number of events in 
the bin.
\label{dt2donfig}}
\end{figure}

\subsection{Fitting the data}

Figure~\ref{dt2donfig} shows a distribution of $\egl$ vs.\ $\egh$ based on 
the on-$\Upsilon(3S)$ data.
To constrain the backgrounds in our fits, we use fit ranges for
$\egl$ and $\egh$ larger than the ranges illustrated
in the figure.
We choose our fit ranges to be $(420<$$\egh$$<560)$~MeV and
$(340<$$\egl$$<460)$~MeV. While the upper (lower) bound of $\egl$($\egh$) does 
not
matter much because of bins containing no events, the choice of the lower
(upper) bound of $\egl$($\egh$) controls 
the statistics available for fitting the
backgrounds. Since the minimum $\egl$ is related to the maximum $\egh$ by the
kinematic constraints, we study the variation of the background scale factor as
a function of the maximum $\egh$.  Based on this exercise, we choose the
maximum $\egh$ to be $560$~MeV and the minimum $\egl$$=340 (\simeq 900-560)$ 
MeV where $\egh+\egl\simeq 900$
values for which the fitted normalization scale factors become stable compared
to the sizes of their statistical errors.

\begin{figure}[hbtp]
\includegraphics[width=0.85\textwidth]{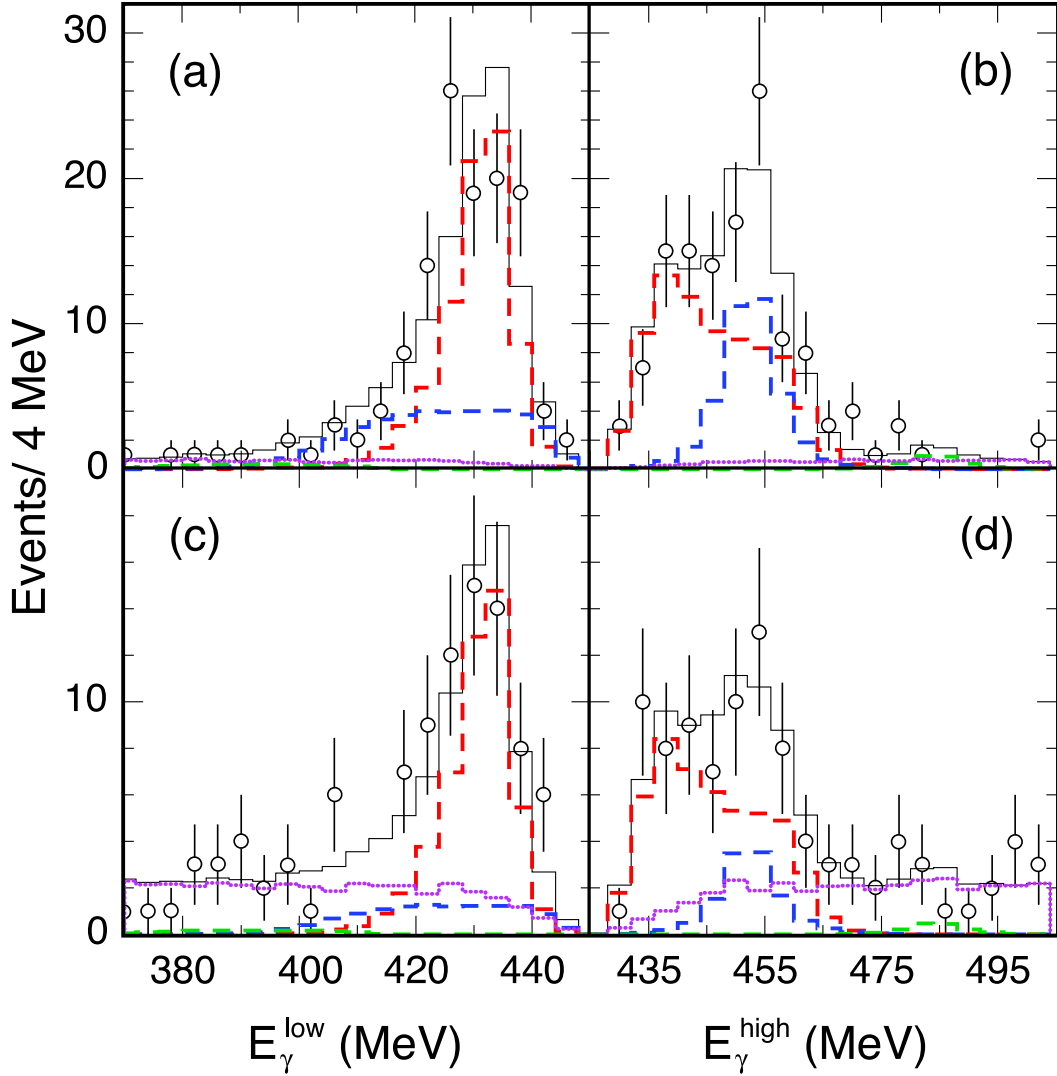}
\caption{Projections from the nominal
fit to data onto (a,c) $\egl$ 
and (b,d) $\egh$ axes for data using the (a,b) $\mumu$ and
(c,d) $\ee$
selections.  The dashed histograms represent signal photons via
$\chi_{bJ}(1P)$.
The dotted histograms represent
the QED MC sample contribution as
normalized by the fit, and the
solid histogram is the sum of
background and signal histograms.\label{datafit}}
\end{figure}

Using the 
QED MC background histograms as the background function
and the three signal
Monte Carlo samples, and fixing the normalization of the
$J=0$ component with the
measured $\brf[\Upsilon(3S)\to\gamma\chi_{b0}(1P)]$~\cite{upsart} and
$\brf[\chi_{b0}(1P)\to\gamma\Upsilon(1S)]$ (this work),
we perform a maximum likelihood fit to the
the 2D data distribution in $\egl$ vs.\
$\egh$.  
We float the normalizations
of both background and signals (for $J=1$ and $2$ only).  
The results of  the fits projected onto the two
photon-energy 
axes are shown in 
Fig.\ \ref{datafit}.

We also extract signal 
yield by subtracting the fitted background
shape, with normalization fixed based on the above nominal 
fit procedure, and then by summing the resultant distribution
over the signal region.
This yield corresponds to the one
from {\it both} transitions, $\Upsilon(3S)\to\gamma\gamma\ell^+\ell^-$ 
via $\chi_{bJ}(1P)$ for $J=1$ and $2$.
The observed yields, along with efficiency-corrected branching fractions,
are shown in Table~\ref{tab:3sfitresult}.

\subsection{Systematic uncertainties~\label{subsec:sys}}

We study the systematic uncertainties in a similar way as in
Sec.\ \ref{subsec:y2ssys}.
We mention only studies whose methods differ from the description
given in Sec.\ \ref{subsec:y2ssys}.
Table~\ref{syserrs} shows a summary of estimated systematic uncertainties.

\subsubsection{Binning scheme}

We calculate the branching fractions using 4 MeV
bins instead of the nominal 2 MeV bins.
Besides a difference in bin sizes, we also consider a difference in bin shapes
with a distribution of
$E_{\rm{SUM}}$ ($=\egl + \egh$) vs.\
$E_{\rm{DIFF}}$ ($=\egh - \egl$).

\subsubsection{ Photon energy resolution}

In order to account for the possibility that the width of the signal MC photon
distributions may not accurately reflect the width of the photon distributions
in data, we have generated additional signal MC sets with the detector
resolution broadened and narrowed by $10\%$ of itself
and recalculated the branching
fractions using these altered MC sets.

\subsubsection{ Photon absolute energy calibration}

To estimate possible systematic effects due to 
miscalibration of the absolute photon energy,
we examine how well the peak values in the kinematically-fitted distributions
of $\egl$ for $\Upsilon(2S)\to\gamma\gamma\Upsilon(1S)$ candidates
are calibrated with respect to those expected.  The largest deviation is
$0.2$~MeV for the $J=2$ peak position
when an $\ee$ final state is selected.
It is well reproduced in our MC samples with input masses of $\Upsilon(1S,2S)$
as well as of $\chi_{bJ}(1P)$ based on the latest information \cite{PDG}.
These photon energies
are kinematically-fitted variables, with $\egl$ and $\egh$
constrained to a fixed value for any given $\Upsilon(1S)$
momentum, so any shift in $\egl$ must be compensated by
a corresponding one of opposite sign in $\egh$.
Based on these observations, we conservatively
vary signal positions by $\egh$$\pm 1$~MeV and
simultaneously by $\egl$$\mp 1$~MeV.

\subsubsection{Additional contributing uncertainties}

Other possible systematic effects that we have
investigated include a variation in the (small) branching fraction
for the transition through $\chi_{b0}(1P)$
and insertion of an explicit fixed
background component from 
$\Upsilon(3S)\to\pi^0\pi^0\Upsilon(1S)$.

\begin{table}[hbtp]

\caption{Estimates of relative systematic uncertainties (in $\%$) for this
analysis, for each $J$, and the sum of $J=1$ and $2$.
The first six entries are calculated for both spins.
The fit to the sum of $J=1$ and $2$ has better stability 
against our binning scheme variations since yields
of $J=1$ and $2$ are statistically anti-correlated, resulting in smaller
variations in terms of their sum.
For the rightmost column,
the entry for ``MC simulation'' includes not only statistical errors of 
reconstruction
efficiencies, but also the total uncertainties of the measured 
branching fractions on the weighted efficiency.
\label{syserrs}}
\begin{center}
\begin{tabular*}{0.65\textwidth}{@{\extracolsep{\fill}}c c c c} \hline \hline
Contribution \Tt \Bb & $J=2$ & $J=1$ & $J=1$ and $2$\\ \hline 
$N_{\Upsilon(3S)}$ \Tt & \multicolumn{3}{c}{$1.7$}\\
 Track-finding & \multicolumn{3}{c}{$1.0$}\\
 Photon-finding & \multicolumn{3}{c}{$2.0$}\\ 
Reduced $\chi^2$ requirement	              & \multicolumn{3}{c}{4.9} \\
Lepton identification 	              & \multicolumn{3}{c}{1.1} \\
Lepton flavor difference \Bb & \multicolumn{3}{c}{1.2} \\
\hline
QED background              \Tt        & ---   & ---   & $8.8$ \\
Include $\upsiii\goesto\pizpiz\upsi$  & $0.3$ & $1.2$ & $0.5$ \\ 
$\chibzero$ Yield variation 	      &	$0.0$ & $0.6$ & $0.2$ \\
Binning scheme                        & $3.3$ & $11.6$& $2.7$ \\
Photon energy resolution              & $2.3$ & $3.7$ & --- \\
Photon absolute energy                & $9.1$ & $11.0$& --- \\ 
MC simulation               \Bb        & ---   & ---   & $0.7$ \\
\hline 
Total                       \Tt \Bb     & $11.6$ & $17.5$ & $11.0$ \\ 
\hline \hline
\end{tabular*}
\end{center}
\end{table}

\subsection{\boldmath Results on analysis of
$\Upsilon(3S)\to\gamma\gamma\ell^+\ell^-$}\label{sec:3sresult}

\begin{table}[htbp]
\caption{Final results of this analysis. Here,
$\brf1 = \brf[\Upsilon(3S)\to\gamma\chi_{bJ}(1P)]$,
$\brf2 = \brf[\chi_{bJ}(1P)\to\gamma\Upsilon(1S)]$, and
$\brf3 = \brf[\Upsilon(1S)\to\ell^+\ell^-]$.
We use $\brf3 = (2.48\pm0.05)\%$~\cite{PDG} and
$\brf2 = (18.5\pm0.5)\%)$, $(33.0\pm0.6)\%$ from 
Sec.\ \ref{sec:2S}
for $J=2$ and $J=1$, respectively, to extract $\brf1\times\brf2$ as well as 
$\brf1$.
The first three rows show results from this work while the second three
rows show previous results.
Here, the first errors are statistical, the second errors are systematic, and
the third errors (when applicable) are uncertainties due to external sources.
}
\label{tab:result3sll}
\begin{center}
\scalebox{0.85}
{
\def\1#1#2#3{\multicolumn{#1}{#2}{#3}}
\begin{tabular*}{1.1\textwidth}{@{\extracolsep{\fill}}c c  c  c } \hline \hline
 \Tt \Bb   &$J=1$      & $J=2$ & $J=1$ and $2$ \\ \hline
 $\brf1\times\brf2\times\brf3~(10^{-5})$ \Tt
    &$1.33\pm0.30\pm0.23$&$3.56\pm0.40\pm0.41$&$4.96\pm0.37\pm0.55$ \\
 $\brf1\times\brf2~(10^{-4})$
    &$5.38\pm1.20\pm0.94\pm0.11$
    &$14.35\pm1.62\pm1.66\pm0.29$&$19.99\pm1.50\pm2.20\pm0.40$ \\
 $\brf1~(10^{-3})$ \Bb
    &$1.63\pm0.36\pm0.28\pm0.09$
    &$7.74\pm0.88\pm0.88\pm0.38$& --- \\
\hline
 $\brf1\times\brf2\times\brf3~(10^{-5})$~\cite{Skwarnicki:2002bp} \Tt \Bb
    & --- & --- & $5.20\pm0.54\pm0.52$ \\
 $\brf1\times\brf2~(10^{-4})$~\cite{cusb} \Tt \Bb
    & --- & --- & $12^{+4}_{-3}\pm0.9$ \\
 $\brf1~(10^{-3})$~\cite{chib} \Tt \Bb
    &$<1.9$
    &$<20.3$ $(11\pm6\pm2\pm1)$ & --- \\
\hline \hline
\end{tabular*}
}
\end{center}
\end{table}

\begin{table}[htb]

\caption{Comparison of measurements and theoretical predictions \cite{thmodels}
for suppressed E1 transition rates $\Gamma[\Upsilon(3S)\to\gamma\chi_{bJ}
(1P)]$ and ratios $\Gamma_{J=1}/ \Gamma_{J=0} \equiv
\Gamma[\Upsilon(3S)\to\gamma\chi_{b1}(1P)]/ \Gamma[\Upsilon(3S)\to\gamma
\chi_{b0}(1P)]$,
$\Gamma_{J=2}/\Gamma_{J=0} \equiv \Gamma[\Upsilon(3S)\to
\gamma\chi_{b2}(1P)]/ \Gamma[\Upsilon(3S)\to\gamma\chi_{b0}(1P)]$,
and $\Gamma_{J=2}/\Gamma_{J=1} \equiv \Gamma[\Upsilon(3S)\to
\gamma\chi_{b2}(1P)]/ \Gamma[\Upsilon(3S)\to\gamma\chi_{b1}(1P)]$.
The CLEO III values are based on
$\Gamma_{\text{total}}[\Upsilon(3S)]=(20.32 \pm 1.85)$ keV \cite{PDG}
and are obtained by taking the
central value of the measurement for the $J=0$ state~\cite{upsart} and
the values for $J=1$ and $2$ from this work.
The last row shows $\Gamma_{J=1}/\Gamma_{J=0}$ and $\Gamma_{J=2}/
\Gamma_{J=0}$ when scaling rates according to $E_\gamma^3\times (2J+1)$.
\label{tab:modelscomp}}
\begin{center}
{
\scalebox{1}
{
\def\1#1#2#3{\multicolumn{#1}{#2}{#3}}
\begin{tabular*}{0.99\textwidth}{@{\extracolsep{\fill}}c c c c c c c} \hline \hline
 \Tt \Bb &$\Gamma_{J=0}$ (eV) & $\Gamma_{J=1}$ (eV) & $\Gamma_{J=1}/\Gamma_{J=0}$
 & $\Gamma_{J=2}$ (eV) & $\Gamma_{J=2}/\Gamma_{J=0}$ & $\Gamma_{J=2}/\Gamma_{J=1}$ \\ \hline
CLEO III (This expt.) \Tt   & -- & $33\pm10$ &$0.54\pm0.25$ & $157\pm30$ & $2.58\pm1.01$ & $4.75\pm1.75$\\
Inclusive expt.~\cite{upsart}       & $61\pm23$ & -- & & -- & & \\
$\chi_{bJ}(1P)$ exclusive expt.~\cite{chib} \Bb & $<186$&$<38$ & & $<413$ & & \\
\hline
Moxhay--Rosner (1983) \Tt & $25$ & $25$ & $1.0$ & $150$ & $6.0$ & $6.0$ \\
Gupta {\it et al.} (1984) & $1.2$ & $3.1$ & $2.6$ & $4.6$ & $3.8$ & $1.5$ \\
Grotch {\it et al.} (1984) (a) & $114$ & $3.4$ & $0.03$ & $194$ & $1.7$ & $57$ \\
Grotch {\it et al.} (1984) (b) & $130$ & $0.3$ & $0.002$ & $430$ & $3.3$ & $1433$ \\
Daghighian--Silverman (1987) & $42$ & (c) &(c) & $130$ & $3.1$ & (c) \\
Fulcher (1990) & $10$ & $20$ & $2.0$ & $30$ & $3.0$ & $1.5$ \\
L\"ahde (2003) & $150$ & $110$ & $0.7$ & $40$ & $0.3$ & $0.4$ \\
Ebert {\it et al.} (2003) \Bb & $27$ & $67$ & $2.5$ & $97$ & $3.6$ & $1.4$ \\ \hline
$E_\gamma^3 \times (2J+1)$ \Tt \Bb & & & $2.4$ & & $3.6$ & $1.5$ \\ \hline \hline
\1{4}{l}{(a) Scalar confining potential.  (b) Vector confining potential.} \\
\1{5}{l}{(c) The authors did not provide a prediction for 
  $\Gamma[\Upsilon(3S)\to\gamma\chi_{b1}(1P)]$.} \\
\end{tabular*}
}
}
\end{center}
\end{table}

Taking the above systematic uncertainties into account, we now arrive at the
final results for the product of branching fractions for each of the
transitions as shown in Table~\ref{tab:result3sll}.  Also shown are comparisons
to results from other analyses.  The first uncertainty is the statistical
uncertainty, the second is the overall systematic uncertainty, and the third
(when applicable) is the uncertainty due to external inputs.

\section{Results and conclusions}\label{sec:result}

We obtain product branching fractions for the exclusive processes
$\Upsilon(2S)\to \gamma \chi_{b0,1,2}(1P) 
\to \gamma \gamma \Upsilon(1S)$ (Table~\ref{tab:result2sll}) and
$\Upsilon(3S)\to \gamma \chi_{b1,2}(1P) 
\to \gamma \gamma \Upsilon(1S)$ (Table~\ref{tab:result3sll}), where
$\Upsilon(1S)$ is identified by its decay to $e^+ e^-$ and $\mu^+ \mu^-$.

The extracted $\brf[\chi_{bJ}(1P) \to \gamma \Upsilon(1S)]$ are
the most precise to date for $J=1,~2$, while for $J=0$ this represents
the first observation of this transition.
These branching fractions appear to be
systematically smaller than the theoretical predictions (see Appendix~\ref{sec:appb}), 
indicating that the hadronic widths of $\chi_{bJ}(1P)$ might have been
underestimated.

\begin{figure}[h]
\begin{center}
\scalebox{1}
{
\includegraphics[width=1\textwidth]{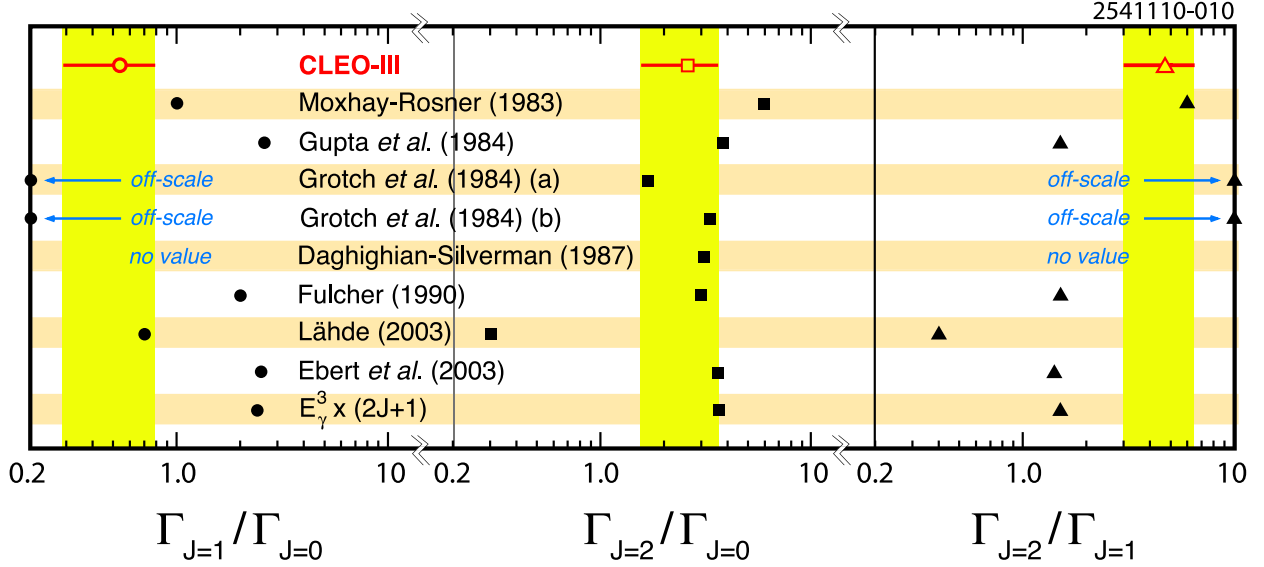}
}
\caption{Illustration of ratios of
suppressed E1 transition rates, $\Gamma_{J=1}/\Gamma_{J=0}$ (circles),
$\Gamma_{J=2}/\Gamma_{J=0}$ (squares), and 
$\Gamma_{J=2}/\Gamma_{J=1}$ (triangles) from Table~\ref{tab:modelscomp}.
\label{fig:rateratio}
}
\end{center}
\end{figure}

The extracted $\brf[\Upsilon(3S)\to\gamma\chi_{b1,2}(1P)]$
may be compared with the branching fraction previously
measured by CLEO, $\brf[\Upsilon(3S) \to \gamma \chi_{b0}(1P)] = (0.30 \pm 0.04
\pm 0.10)\%$~\cite{upsart},
providing tests of relativistic corrections to electric dipole
matrix elements. 
Table~\ref{tab:modelscomp} shows comparison against some
theoretical predictions in terms of transition rates
as well as ratios of transition rates while
Fig.\ \ref{fig:rateratio} shows the ratios pictorially.
It might be worth revisiting these calculations in light of our new 
experimental results.

\begin{acknowledgments}
We gratefully acknowledge the effort of the CESR staff in providing us with
excellent luminosity and running conditions.  D.~Cronin-Hennessy thanks the
A.P.~Sloan Foundation.  J. Rosner thanks Fermilab for hospitality during
part of this investigation.  This work was supported by the National Science
Foundation, the U.S. Department of Energy, the Natural Sciences and Engineering
Research Council of Canada, and the U.K. Science and Technology Facilities
Council.
\end{acknowledgments}

\appendix

\section{\boldmath Comparison of branching fractions 
$\brf[\chi_{bJ}(1P)\to\gamma\Upsilon(1S)]$ with predictions}\label{sec:appb}

The measured branching fractions for $\chi_{bJ}(1P) \to \gamma \Upsilon(1S)$
may be compared with the predictions \cite{thmodels,KR} summarized in Table
\ref{tab:comp}.  
Most of the predicted branching fractions for these electric dipole
transitions are systematically larger than the experimental values, indicating
that the hadronic widths $\Gamma_h$ were underestimated.  A modest increase in
the assumed value of $\alpha_S(m_b^2)$ leads to much better agreement with
experiment.  As one example, the values in Ref.\ \cite{KR} were calculated
for $\alpha_S(m_b^2) = 0.18$.  For this value it was found that
$\Gamma_h[\chi_{b(0,1,2)}(1P)] = (791,38.3,132.3)$ keV, while the E1 transition
rates were predicted to be $\Gamma[\chi_{b(0,1,2)}(1P) \to \gamma \Upsilon(1S)]
= (26.1,32.8,37.8)$ keV.  The hadronic widths scale for the $J=0$ and $J=2$
states $\alpha^2_S(m_b^2)$ times known QCD correction factors \cite{KMRR},
while for the $J=1$ state they scale as $\alpha^3_S(m_b^2)$.  (The QCD
correction factor for $J=1$ is not known \cite{KMRR} and will be ignored.)

\begin{table}
\caption{Comparison of our results for ${\cal B}[\chi_{bJ}(1P) \to \gamma
\Upsilon(1S)]$ with some theoretical predictions \cite{thmodels,KR},
in units of $10^{-2}$.
\label{tab:comp}}
\begin{center}
\scalebox{1}
{
\def\1#1#2#3{\multicolumn{#1}{#2}{#3}}
\begin{tabular*}{0.8\textwidth}{@{\extracolsep{\fill}}c c c c} \hline \hline
Reference \Tt \Bb & $J=0$ & $J=1$ & $J=2$ \\ \hline
CLEO III \Tt & $1.73\pm0.35$ & $33.0\pm2.6$ & $18.3\pm1.4$ \\
Moxhay--Rosner (1983)  & 3.8 & 50.6 & 22.3 \\
Gupta {\it et al.} (1984) & 4.1 & 56.8 & 26.7 \\
Grotch {\it et al.} (1984) (a) & 3.1 & 41.9 & 19.4 \\
Grotch {\it et al.} (1984) (b) & 3.3 & 43.9 & 20.3 \\
Daghighian--Silverman (1987) & 2.3 & 31.6 & 16.6 \\
Kwong--Rosner (1988) & 3.2 & 46.1 & 22.2 \\ 
Fulcher (1990) & 3.1 & 39.9 & 18.6 \\
L\"ahde (2003) & 3.3 & 45.7 & 21.1 \\
Ebert {\it et al.} (2003) & 3.7 & 51.5 & 23.6 \\
\hline \hline
\1{4}{l}{(a) Scalar confining potential.  (b) Vector confining potential.} \\
\end{tabular*}
}
\end{center}
\end{table}

Using the scale factors, values of $\alpha_S$, and the QCD correction
factors described in the previous paragraph, we predict
\bea
\frac{\Gamma_h[\chi_{b0}(1P)]}{791~{\rm keV}}&=&\left( \frac{\alpha_S(m_b^2)}
{0.18} \right)^2 \frac{1 + 10.0 \alpha_S(m_b^2)/\pi}{1.573}, \nonumber \\
\frac{\Gamma_h[\chi_{b1}(1P)]}{38.3~{\rm keV}}&=&\left( \frac{\alpha_S(m_b^2)}
{0.18} \right)^3,\ \mathrm{and} \nonumber \\
\frac{\Gamma_h[\chi_{b2}(1P)]}{132.3~{\rm keV}}&=&\left( \frac{\alpha_S(m_b^2)}
{0.18} \right)^2 \frac{1 - 0.1 \alpha_S(m_b^2)/\pi}{0.994}~.
\eea
The minimum $\chi^2$ of a fit to the values determined in this work is found
to be 1.45 for 2 
degrees of freedom
The value of $\alpha_S$ that minimizes $\chi^2$ and the corresponding
error (defined by the range for which 
$\Delta\chi^2\le 1$ from the minimum) are
$\alpha_S(m_b^2) = 0.214 \pm 0.006$.  This is
quite consistent with the determination of Ref.\ \cite{Bethke:2009jm} for a
scale of 5 GeV (see Fig.\ 5 there).  At this value, the rescaled values
predicted in the approach of Ref.\ \cite{KR} are  ${\cal B} [\chi_{b0,1,2}(1P)
\to \gamma \Upsilon(1S)] = (2.1,33.8,16.8)\%$ and $\Gamma_{\rm tot}
[\chi_{b0,1,2}(1P)] = (1221,97,225)$ keV.

The predicted ratio
\beq
R \equiv \frac{\Gamma_h(\chi_{b0}(1P))}{\Gamma_h(\chi_{b2}(1P))} = \frac{15}{4}
\frac{1 + 10.0 \alpha_S(m_b^2)/\pi}{1 - 0.1 \alpha_S(m_b^2)/\pi}
\eeq
is roughly $R = 5.92 + 12[\alpha_S(m_b^2) - 0.18]$, to be compared with the
value $R = 8.6 \pm 3.2$ based on the observed branching fractions.  Thus,
the QCD corrections go in the right direction to modify the uncorrected
value of 15/4 = 3.75.


\begin{thebibliography}{99}

\bibitem{Eichten} E. Eichten {\it et al.}, Rev.\ Mod.\ Phys.\ {\bf 80}, 1161
(2008).

\bibitem{PDG} K. Nakamura {\it et al.} (Particle Data Group), J. Phys.\ G
{\bf 37}, 075021 (2010).

\bibitem{Grant:1995hf}
  A.~K.~Grant, J.~L.~Rosner, A.~Martin, J.~M.~Richard and J.~Stubbe,
  Phys.\ Rev.\ D {\bf 53}, 2742 (1996)
  [arXiv:hep-ph/9506315].

\bibitem{thmodels}  P. Moxhay and J. L. Rosner, Phys.\ Rev.\ D {\bf 28}, 1132
(1983); S. N. Gupta, S. F. Radford, and W. W. Repko, Phys.\ Rev.\ D
{\bf 30}, 2424 (1984); H. Grotch, D. A. Owen, and K. J. Sebastian,
Phys.\ Rev.\ D {\bf 30}, 1924 (1984); F. Daghighian and D. Silverman,
Phys.\ Rev.\ D {\bf 36}, 3401 (1987); J. P. Fulcher,
Phys.\ Rev.\ D {\bf 42}, 2337 (1990); T. A. L\"ahde,
Nucl.\ Phys.\ A {\bf 714}, 183 (2003); D. Ebert, R. N. Faustov, and
V. O. Galkin, Phys.\ Rev.\ D {\bf 67}, 014027 (2003).

\bibitem{Heintz:1992}
U. Heintz {\it et al.} (CUSB Collaboration), Phys.\ Rev.\ D {\bf 46}, 1928
(1992).

\bibitem{Skwarnicki:2002bp} T.~Skwarnicki (CLEO Collaboration),
in {\it Proceedings of the 31st International Conference on High Energy Physics (ICHEP 2002), Amsterdam, The Netherlands, 24-31 Jul 2002}, 
edited by S. Bentvelsen, P. de Jong, J. Koch, and E.Laenen 
[Nucl.\ Phys.\ B, Proc. Suppl. 117, 698 (2003)],
available online by inputting the author’s name (Skwarnicki) and the journal 
title (unabbreviated) at http://www.sciencedirect.com.

\bibitem{chib} D. M. Asner {\it et al.} (CLEO Collaboration),
Phys.\ Rev.\ D {\bf 78}, 091103 (2008).

\bibitem{etaart} Q. He {\it et al.} (CLEO Collaboration), Phys.\ Rev.\ Lett.\
{\bf 101}, 192001 (2008).

\bibitem{upsart}
M. Artuso {\it et al.} (CLEO Collaboration),
Phys.\ Rev.\ Lett.\ {\bf 94}, 032001 (2005).

\bibitem{CLEO2}
Y.~Kubota {\it et al.},
Nucl.\ Instrum.\ Meth.\ A {\bf 320}, 66 (1992).

\bibitem{CLEO3trk} D.~Peterson {\it et al.},
Nucl.\ Instrum.\ Meth.\ A {\bf 478}, 142 (2002).

\bibitem{CLEOcuts} J.~V.~Bennett {\it et al.} (CLEO Collaboration), Phys.\
Rev.\ Lett.\ {\bf 101}, 151801 (2008); R. E. Mitchell {\it et al.} (CLEO
Collaboration), Phys.\ Rev.\ Lett.\ {\bf 102}, 011801 (2009); P. U. E. Onyisi
{\it et al.} (CLEO Collaboration), Phys.\ Rev.\ D {\bf 82}, 011103(R) (2010).

\bibitem{evtgen} D.J. Lange, Nucl. Instrum. Methods Phys. Res., Sect. A
{\bf 462}, 152 (2001).

\bibitem{pipiart} S. R. Bhari {\it et al.} (CLEO Collaboration),
Phys.\ Rev.\ D {\bf 79}, 011103 (2009).

\bibitem{cbal} W. Walk {\it et al.} (Crystal Ball Collaboration),
Phys.\ Rev.\ D {\bf 34}, 2611 (1986).

\bibitem{cusb0} F. Pauss {\it et al.} (CUSB Collaboration),
Phys.\ Lett.\ B {\bf 130}, 439 (1983).

\bibitem{cusbj} C. Klopfenstein {\it et al.} (CUSB Collaboration),
Phys.\ Rev.\ Lett.\ {\bf 51}, 160 (1983).

\bibitem{babayaga} C. M. Carloni Calame {\it et al.},
Nucl.\ Phys.\ Proc.\ Suppl.\ {\bf 131}, 48 (2004);
C. M. Carloni Calame, Phys.\ Lett.\ B {\bf 520}, 16 (2001);
C. M. Carloni Calame {\it et al.}, Nucl.\ Phys.\ B {\bf 584}, 459 (2000).

\bibitem{cusb} U. Heintz {\it et al.} (CUSB Collaboration),
Phys.\ Rev.\ D {\bf 46}, 1928 (1992).

\bibitem{KR} W. Kwong and J. L. Rosner, Phys.\ Rev.\ D {\bf 38}, 279 (1988).

\bibitem{KMRR} W. Kwong. P. Mackenzie, R. Rosenfeld, and J. L. Rosner,
Phys.\ Rev.\ D {\bf 37}, 3210 (1988).

\bibitem{Bethke:2009jm}
S.~Bethke,
  Eur.\ Phys.\ J.\  C {\bf 64}, 689 (2009)
  [arXiv:0908.1135 [hep-ph]].

\end{thebibliography}
\end{document}